\documentclass{article}

\usepackage{arxiv}

\usepackage[utf8]{inputenc} 
\usepackage[T1]{fontenc}    
\usepackage{amsmath}
\usepackage{hyperref}       
\usepackage{url}            
\usepackage{booktabs}       
\usepackage{amsfonts}       
\usepackage{nicefrac}       
\usepackage{microtype}      
\usepackage{cleveref}       
\usepackage{lipsum}         
\usepackage{graphicx}
\usepackage{natbib}
\usepackage{doi}
\usepackage{dcolumn}

\title{The origin of the terrestrial planets}

\author{ {R. B.~Firestone} \\
	Lawrence Berkeley National Laboratory\\
	Department of Nuclear Engineering\\
	University of California, Berkeley 94720\\
	\texttt{rbfirestone@lbl.gov} \\
}

\renewcommand{\shorttitle}{\textit{arXiv} Template}

\begin{document}
\maketitle

\begin{abstract}

Three major planets, Venus, Earth, and Mercury formed out of the solar nebula.  A fourth planetesimal, Theia, also formed near Earth where it collided in a giant impact, rebounding as the planet Mars.  During this impact Earth lost ${\approx}4$\% of its crust and mantle that is now is found on Mars and the Moon.  At the antipode of the giant impact, $\approx$60\% of Earth's crust, atmosphere, and a large amount of mantle were ejected into space forming the Moon.  The lost crust never reformed and became the Earth's ocean basins.  The Theia impact site corresponds to Indian Ocean gravitational anomaly on Earth and the Hellas basin on Mars.  The dynamics of the giant impact are consistent with the rotational rates and axial tilts of both Earth and Mars.  The giant impact removed sufficient CO$_2$ from Earth's atmosphere to avoid a runaway greenhouse effect, initiated plate tectonics, and gave life time to form near geothermal vents at the continental margins.  Mercury formed near Venus where on a close approach it was slingshot into the Sun's convective zone losing 94\% of its mass, much of which remains there today.  Black carbon, from CO$_2$ decomposed by the intense heat, is still found on the surface of Mercury.  Arriving at 616 km/s, Mercury dramatically altered the Sun's rotational energy, explaining both its anomalously slow rotation rate and axial tilt.  These results are quantitatively supported by mass balances, the current locations of the terrestrial planets, and the orientations of their major orbital axes.

\end{abstract}

\section{Introduction}

The solar system formed from a dense cloud of hydrogen and helium rich interstellar gas and dust that collapsed due to the shockwave of a nearby supernova that formed the solar nebula, a spinning, swirling disk.  Gravitational forces collapsed the nebula increasing its rotational velocity, much like a ballerina who pulls in her arms to rotate her pirouette faster.  Centrifugal forces would have separated the heavier elements away from the center much like a gas centrifuge.  As the rotational velocity reached Mach$\approx$4-5 (1200-1500 m/s) fusion occurred~\cite{Ellis2001} and the Sun was born.  Simultaneously, far from the center of the solar nebula where the temperature was cold enough, volatiles condensed forming the cores of Jupiter and the outer planets gathering hydrogen and helium from their surroundings to becoming gas giants.

Very little hydrogen or helium remains between the Sun and Jupiter indicating that centrifugal forces and the formation of the gas giants had completely separated the heavy elements from the light gases, possibly even before the Sun had ignited.  The heavy elements began to aggregate near Earth's orbit into dust particles that continued to collide forming rocks, protoplanets, and ultimately the terrestrial planets.  The terrestrial planets all would have formed within a narrow distance from the Sun, initially circulating in spherical orbits with similar compositions.  When only a few large terrestrial planets and protoplanets remained, their final collisions created the inner solar system that we see today.

The compositions of the terrestrial planets no longer resemble their meteoritic origins.  As they formed the lighter elements rose to the surface as less dense crust and mantle while iron and nickel sank into the denser core.  A collision between the Earth and the protoplanet Theia removed crust and mantle forming the less dense Mars and the Moon.  Conversely Mercury, which is an iron rich planet, lost most of its crust, mantle, and core during a close encounter with the Sun's convection zone.  In this work I will present evidence that these events can be accounted for by mass balances, geographical evidence, and the motions of the planets dictated by the laws of conservation of energy and momentum.

\begin{table*}
\centering
\caption{Composition of the  terrestrial planets and Earth's Moon.}
\label{Terrestrial}
\begin{tabular}{llllll}
\toprule
&Earth&Venus&Mercury&Mars&Moon\\
\midrule
Eccentricity&0.0167&0.0068&0.2056&0.0934&0.0549\\
Obliquity (degrees)&23.44&2.64&0.03&25.19&6.68\\
Distance from Sun (AU)&1.000&0.722&0.387&1.52&0.0026$^a$\\
Mass ($10^{10}$ Gt)&597.22(6)&486.75(5)&33.011(3)&64.171(6)&7.342(7)\\
Radius (km)&6378.1&6051.8(10)&2439.7(10)&3389.5(2)&1738.1\\
Radius core (km)&3485(3)&3147(17)&2011(180)&1830(40)&381(12)\\
Iron (wt\%)$^b$&32.07&31.17&68.47&23.7&$9\pm4$\\
Density (g/cm$^3$)&5.51&5.24&5.43&3.93&3.34\\
\bottomrule
\multicolumn{6}{l}{$^a$ Mean distance from Earth, $^b$ \cite{Morgan80}, \cite{Yoshizaki20}, }\\
\multicolumn{6}{l}{\cite{Parkin73}.}\\
\end{tabular}
\end{table*}

\section{The terrestrial planets}

Today, the four terrestrial planets, Mercury, Venus, Earth, Mars, and Earth's Moon remain.  Their compositions are summarized in Table~\ref{Terrestrial}.  Both Earth and Venus have similar, nearly spherical orbits, comparable masses, similar densities, comparable iron abundances, and meteoritic compositions.  They represent over 90 wt\% of the terrestrial planet mass and can be considered as the primary planets whose composition has changed little since their earliest formation.  The planets Mercury and Mars have more elliptical orbits, much smaller masses, and very different iron abundances.  If Mercury and Mars formed near Earth and Venus they would originally have had the same compositions as Earth and Venus.  Their elliptical orbits suggest that they were scattered from their original orbits to their current positions, and their different compositions suggest that they have lost or gained mass due to collisions after their initial formation.  The Moon is unique to Earth with a low density and little or any core indicating that it was never a protoplanet and more likely formed out of the Earth.  Mars, with its low density and small core, is consistent with a protoplanet that accreted additional crust and mantle during a collision.  Mercury, with its massive iron core, appears to have lost much of its crust and mantle during a catastrophic event.

\section{The Giant-Impact hypothesis}

The Giant-Impact hypothesis~\cite{Hartmann75,Canup01} proposes that a Mars-sized object struck the Earth with a velocity of 9.3 km/s, at an oblique angle, producing ejecta that coalesced forming the Moon.  It is supported by the similar orientation of the orbits of the Earth and the Moon, the high angular momentum of the Earth-Moon system, evidence that the Moon was once molten down to great depth, the Moon's tiny core, and the same oxygen isotope composition on the Moon as on Earth.  Arguments against the hypothesis include the low dynamic probability that the impact of a large protoplanet would produce the momentum and obliquity of the Earth-Moon system~\cite{Boss86}, the lack of evidence that the Earth's surface experienced significant melting~\cite{Ringwood89}, and the fact than only one significant moon exists in the inner solar system.  The low density of the Moon, 3.34 g/cm$^3$, compared to that of Earth, 5.51 g/cm$^3$, rules out the capture of a protoplanet.  If a protoplanet formed near Earth's orbit, it would have a composition comparable to Earth.  It is apparent that the Moon must have formed from the crust and mantle of the Earth and/or an impacting protoplanet in order to explain its low density.

The dynamics of how the giant impact produced the Moon remain debatable.  An early argument, suggested by George Darwin~\cite{Darwin1898}, is that if the Earth were spinning rapidly with 3-4 hour day, crust and mantle could be flung off of the surface forming the Moon.  This model fails to explain how the Earth initially rotated so rapidly.  More recently, a similar model Synestia has been proposed~\cite{Lock18} suggesting that a protoplanet called Theia impacted Earth causing it to spin rapidly with a 2-3 hr day~\cite{Cuk12} vaporizing it, and spinning mass off into space that ultimately coalesced forming the Moon.  In order for the giant impact to induce such a high rotational rate either Theia was already rotating extremely rapidly or the impact induced a large torque on Earth during the impact.  If the impact vaporized both Theia and the Earth there is no reason to assume that significant rotation of the dust cloud would happen.  If Earth's day were 3 hr, its rotational energy would be $1.36\times10^{31}$ J, $64\times$ that of Earth's current rotational energy and half of the total giant impact energy.  Conservation of energy requires that the total energy of the Earth-Moon system should not have changed since the giant impact, yet this implies that the Moon should orbit 1100 km from Earth today.  Clearly the hypothesis that the Moon spun off of a rapidly spinning Earth seems unsustainable.

\subsection{The giant rebound}

\begin{figure*}
  \centering
  \includegraphics[width=0.9\textwidth]{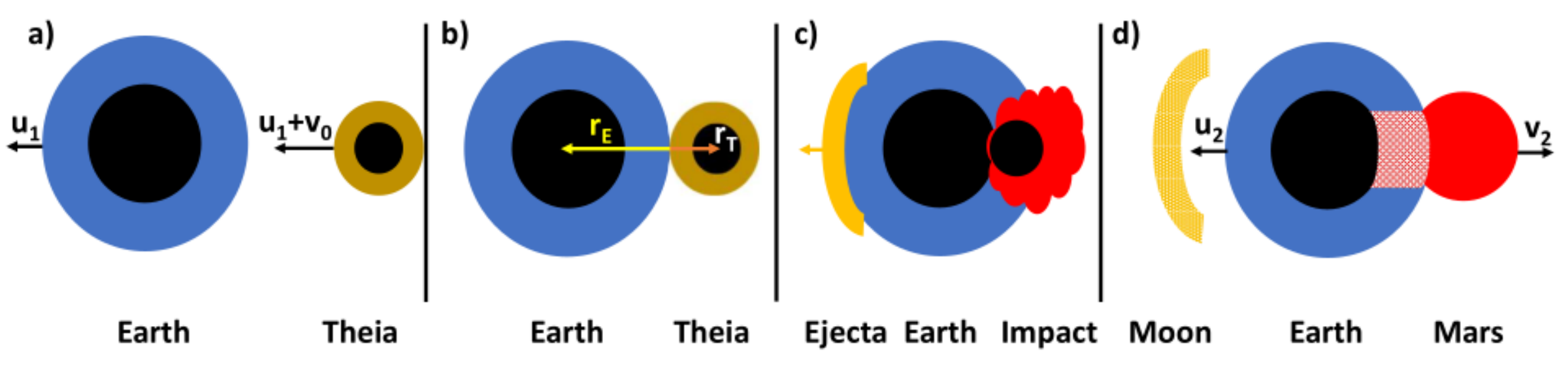}
  \caption{The protoplanet Theia approaches Earth (a) where it collides at a distance $r_E+r_T$ (b) where $r_E$ and $r_T$ are the radii of Earth and Theia respectively.  The impact melts and vaporizes Earth's crust and mantle near the impact site and ejects crust and mantle at the antipode (c).  The planet Mars rebounds from the impact site and a ring forms from the antipode ejecta (d).}
  \label{EMM}
\end{figure*}

\begin{figure}
  \centering
\includegraphics[width=0.4\textwidth]{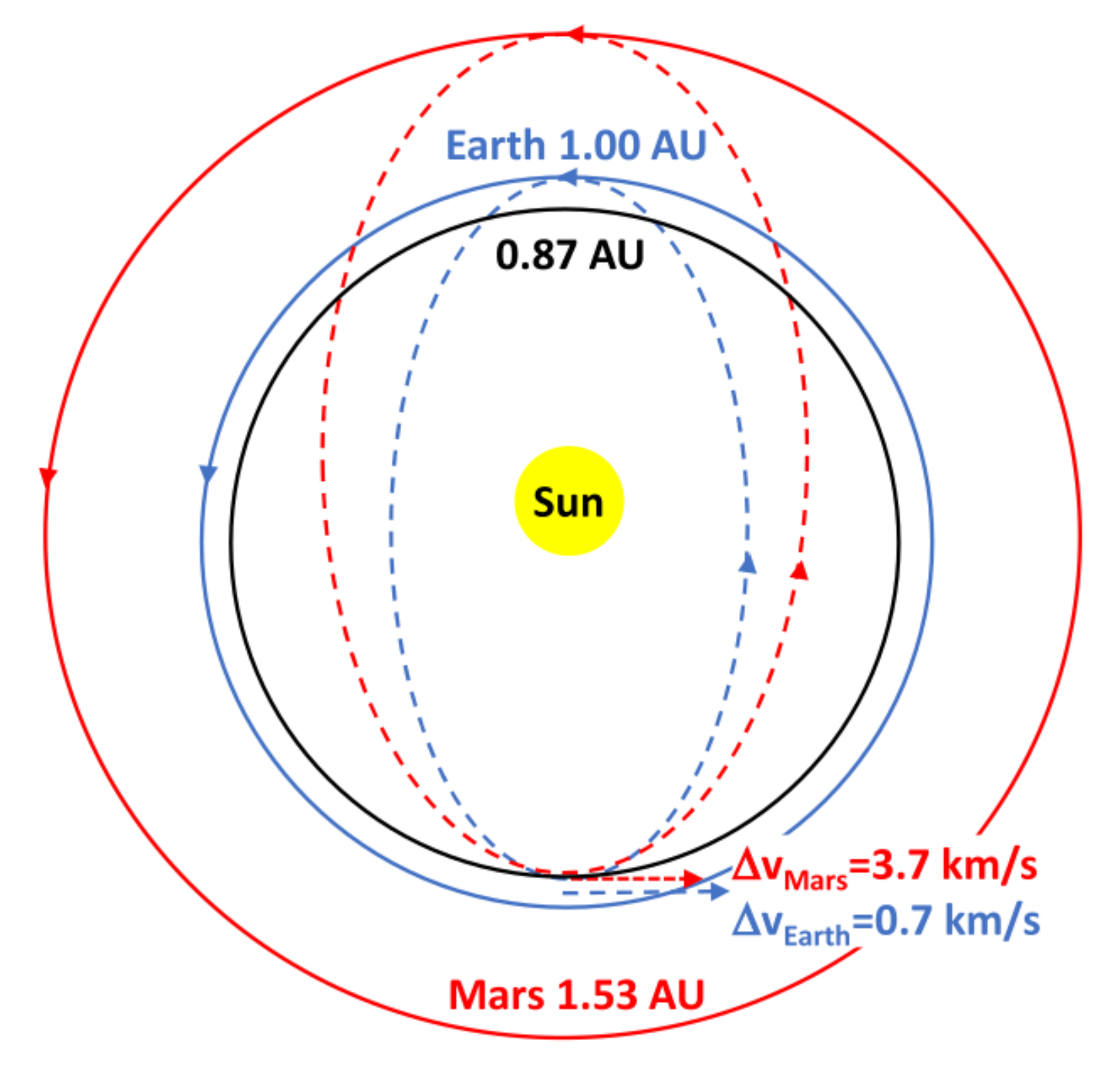}
  \caption{Hohmann transfer orbits from the Earth to the Sun and Mars.}
  \label{Horbit}
\end{figure}

The assumption that a collision of Theia with Earth would have vaporized the planet is deeply flawed.  It implies that Earth was fixed in space so that the entire impact energy was converted into heat.  That fails to conserve the momentum of the Theia/Earth system because both bodies were freely traveling in their orbits before the collision.  A better description of the impact would be the collision of two solid spheres moving through space as shown in Fig.~\ref{EMM}.
Theia, initially traveling at nearly the same orbital velocity, $u_1$, as Earth, would have gained an additional velocity, $v_0$, given by Eq.~\ref{V1}, from gravitational attraction
\begin{equation}\label{V1}
v_0 = \sqrt{2GM_E/(r_E+r_T)}
\end{equation}
when it contacted Earth where  $G=6.67\times10^{-11}~m^3kg^{-1}s^2$is the gravitational constant, $M_E$ is the mass of the Earth, and $r_E$, and $r_T$ are the radii of Earth and Theia respectively.  At impact, Theia and a large amount of Earth's crust and mantle near the impact site would begin to melt and vaporize.  Since the Earth is nearly incompressible, pressure would build forming a reverse shockwave that would force a large volume of crust and mantle to rebound from the impact site with a velocity $v_2$, as the Earth continued forward with a velocity $u_2$.  The rebounding mass would undergo a Hohmann transfer, as shown in Fig.~\ref{Horbit}, moving it into the orbit of Mars, while Earth would move to its current orbit.

\begin{figure}
  \centering
  \includegraphics[width=0.47\textwidth]{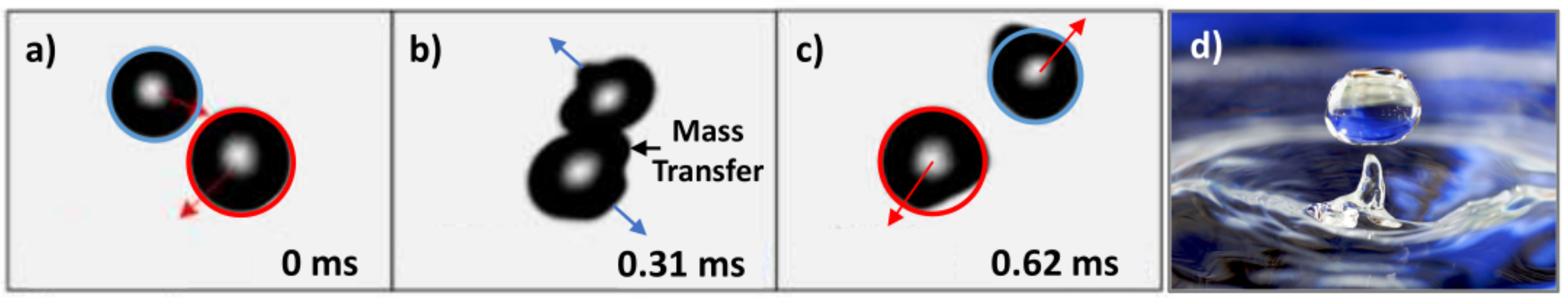}
  \caption{Photo of collision between liquid drops~\protect\cite{Islamova22}.  The drops approach obliquely (a).  At the impact point each drop bulges at its antipode (black arrows) and mass is exchanged between the drops (b).  As the drops move apart the upper drop (blue circle) has gained mass while the lower drop (red circle) has lost mass (c). A drop rebounding from a liquid surface (d) shows a clear antipodal rise with evidence of a Rayleigh instability as the drop pulls away from the surface (d).}
  \label{Drop}
\end{figure}

An analogy to the giant impact has been observed in the collision of liquid drops as shown in Fig.~\ref{Drop}
where the drops scatter while exchanging mass.

\subsection{The antipode mass ejection}

The giant impact produced a shockwave resonating through Earth and amplified at the antipode of the impact site.   This phenomenon was observed in the Chicxulub impact where volcanism occurred near the antipode~\cite{Hagstrum05}.  It is estimated that Earth's surface was displaced by ${\approx}$4 m~\cite{Meschede11}.  The Chicxulub impact released between $1.3{\times}10^{24}$ J and $5.8{\times}10^{25}$ J~\cite{Durand14}, yet the Theia impact would have provided ${\approx}2{\times}10^{31}$ J.  Assuming that the antipode displacement varies linearly with impact energy, this suggests that a giant impact would have displaced Earth's surface by 1400-61000 km, launching it into orbit.  The giant impact would have created a dust ring around Earth that ultimately must have accreted into the Moon.

The mass ejection from the giant impact at its antipode on Earth should be mirrored by a similar mass ejection at the antipode of the impact on Theia that should have left its mark on Mars.  It has been proposed that a giant impact on Mars, near the time the Moon formed, created a broad expanse of lowlands and ejected $10^8$ Gt of debris~\cite{Hesselbrock17}.  Crust and mantle would have been ejected with high velocity from Mars, due to its weaker gravity, and it would likely have escaped into the asteroid belt.

\subsection{Mars crustal dichotomy}

\begin{figure*}
  \centering
\includegraphics[width=\textwidth]{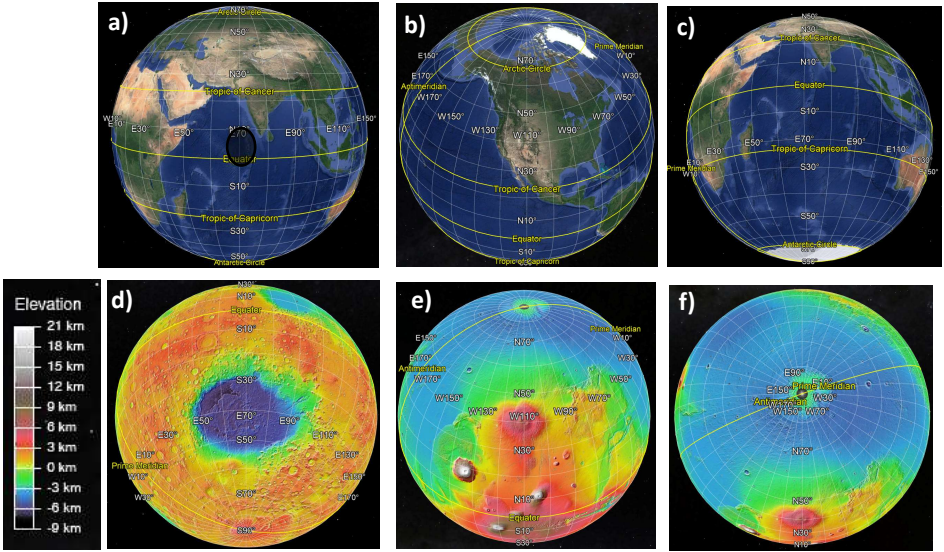}
\vspace{-3cm}
\caption{Google Earth view of the giant impact sites.  a) Indian Ocean gravitational anomaly ($5^{\circ}N,72^{\circ}E$), b) antipode to the Indian Ocean gravitational anomaly, c) antipode to the Indian Ocean gravitational anomaly shifted $25^{\circ}$ north, d) Mars Hellas Basin ($42^{\circ}S,70^{\circ}E$), e)antipode to Mars Hellas Basin, f) Mars Hellas Basin shifted $40^{\circ}$ south.}
\label{MarsCD}
\end{figure*}

Mars has a crustal dichotomy where the Norther hemisphere has a continuous thin crust, averaging 34 km thick, and the Southern Hemisphere has a continuous thick crust averaging 58 km thick.  It is estimated that Mars was once covered by the same fraction of its surface with water as Earth is today~\cite{Saberi20}.  This dichotomy is ascribed to a large impact centered on the Hellas Basin ($42^{\circ}S,70^{\circ}E$).  If the impact were the giant impact, then a similar impact region should be found on Earth.  A candidate for that would be the Indian Ocean gravitational anomaly ($5^{\circ}N,72^{\circ}E$), a large region of low gravity often ascribed to a mantle plume.  This would be consistent with the giant impact penetrating deep into Earth's mantle as it transferred mass from Earth to Mars.

Google Earth views of Mars, centered on Hellas Basin, and on Earth, centered on the Indian Ocean gravitational anomaly, are shown in Fig.~\ref{MarsCD}.  Nearly all of the terrain surrounding Hellas Basin is highlands while most of the Earth's land mass is in the same hemisphere as the Indian Ocean gravitational anomaly.  The views of the antipodes to Hellas Basin and the Indian Ocean gravitational anomaly are also shown in Fig.~\ref{MarsCD}.  The antipode of the Indian Ocean gravitational anomaly is in North America, and the antipode of Hellas Basin is at the edge of Mars highlands.  The Mars lowlands and the Earth's oceans should be displaced from the antipode of the impact site if the impact angle was oblique to the center of the planet.  In Fig.~\ref{MarsCD} shifting the Hellas Basin antipode by $42^{\circ}$ South puts it the center of the lowlands, while shifting the antipode of the Indian Ocean gravitational anomaly by $25^{\circ}$ north puts it in the middle of the ocean.  This indicates that plate tectonics have made only small incursions into Earth's oceans where the giant impact removed crust and mantle.  Although the volcano Alba Mons lies at the exact antipode of Hellas Basin, it actually is at the edge of Mars lowlands near many other Martian volcanoes.  This is analogous to Earth where volcanic activity is focused towards the ring of fire near the subduction zones beneath the continental crust.  It suggests that early Mars may have had plate tectonics.

\subsection{The composition of Mars}

The composition of Mars is very different from that of Earth or Venus and inconsistent with having been formed near either planet.  Its density is 3.93 g/cm$^3$, far less that Earth's, 5.51 g/cm$^3$, but more comparable to the Moon, 3.34 g/cm$^3$.  The abundance of iron on Mars is 23.7 wt\%~\cite{Yoshizaki20}, which is much lower that of Earth, 32.1 wt\%, or Venus, 31.2 wt\%~\cite{Morgan80}.  This is consistent with Theia accreting a large amount of crust and mantle from the Earth as it rebounded from the giant impact.  If Theia initially had the same iron abundance as Venus and it removed crust and mantle from Earth with a 6 wt\% iron abundance, then Mars, with a current mass of $6.42{\times}10^{11}$ Gt, would be composed of Theia, whose mass was $4.5\times10^{11}$ Gt, plus $1.9\times10^{11}$ Gt of mantle and crust that was transferred from the Earth.  This would mean that Earth lost ${\approx}$3\% of its primordial crust and mantle from the impact, a quantity that needs to be accounted for.  As Mars recoiled from Earth it would have been a molten mixture of crust, mantel, and core.  Cooling rapidly its gravity may have been too weak to fully separate its composition into well defined core, mantle, and crust, as occurred on Earth and Venus.  This implies that Mars' surface would have an Earth-like mantle composition with an excess of iron that failed to sink into the core.  The average composition of Mars surface~\cite{Yoshizaki20} is compared to the composition of Earth's continental crust~\cite{CRC97} and upper mantle~\cite{Hart86,McDonough95,Palme03,Lyubetskaya07,Wood96} in Table~\ref{MarsComp}.  As predicted, the composition of Mars' surface is nearly identical to Earth's mantle with an iron abundance that is double that of either Earth's crust or mantle.  Mars elemental composition is consistent with a mixture of Earth's and Theia's crust and mantle and Theia's core.

\begin{table}
  \centering
\tabcolsep=6pt
  \caption{Comparison of the primary elemental composition of Mars surface~\protect\cite{Yoshizaki20} with Earth's continental crust~\protect\cite{CRC97} and upper mantle~\protect\cite{Hart86,McDonough95,Palme03,Lyubetskaya07,Wood96}.}
  \label{MarsComp}
\begin{tabular}{rccccc}
\toprule
&Mars (wt\%)&\multicolumn{2}{c}{Earth(wt\%)}&\multicolumn{2}{c}{Ratio}\\
&Surface&Mantle&Crust&Mantle/Mars&Crust/Mars\\
\midrule
O&43.0&44.7&46.1&1.04&1.07\\
Si &21.3&21.2&28.2&1.00&1.33\\
Mg &18.7&22.3&2.33&1.19&0.12\\
Ca&2.06&2.45&4.15&1.19&2.02\\
Al &1.90&2.15&8.23&1.13&4.33\\
K&0.036&0.023&2.09&0.64&58.55\\
Ti&0.10&0.12&0.57&1.18&5.55\\
Fe&11.4&6.3&5.6&0.55&0.49\\
\bottomrule
\end{tabular}
\end{table}

\begin{table}
  \centering
  \small
\tabcolsep=2pt
  \caption{Comparison of the compositions of Earth and Venus.  The densities of Earth's crust, mantle, and core are calculated from their reported masses and the core and crustal radii.  The masses of Venus' crust, mantle, and core are calculated assuming the same densities as on Earth.  The Earth's crust and mantle mass is depleted by 5.6\% with respect to Venus.}
  \label{EaVe}
\begin{tabular}{lll}
&Earth&Venus\\
\midrule
Mass (Mt)&$5.9722(6){\times}10^{12}$&$4.8675(5){\times}10^{12}$\\
Radius (km)&6378.1&6051.8(10)\\
Radius Crust (km)&&20\\
Mass Crust Mt)&$2.77{\times}10^{10}$$^b$&$2.6{\times}10^{10}$\\
Density Crust (g/cm$^3$)&&2.83$^c$\\
Mass Mantle (Mt)&$4.01(4){\times}10^{12}$$^d$&$3.54(6){\times}10^{12}$\\
Density Mantle (g/cm$^3$)&4.47(4)&\\
Radius Core (km)&3483(5)&3147.1$\pm$16.6\\
Mass Core (Mt)&$1.93(5){\times}10^{12}$$^e$&$1.43(3){\times}10^{12}$\\
Density Core (g/cm$^3$)&10.96(11)&\\
Core+Mantle+Crust (Mt)&$5.97{\times}10^{12}$&$4.99(13){\times}10^{12}$\\
(Mantle+Crust)/Total&0.676(3)&0.714(13)\\
\bottomrule
\multicolumn{3}{l}{$^a$\cite{Maia22}, $^b$\cite{Peterson07}, }\\
\multicolumn{3}{l}{$^c$\cite{Christensen95}, $^d$\cite{Lodders98}, }\\
\multicolumn{3}{l}{$^e$ Calculated from the difference between Earth's total mass}\\
\multicolumn{3}{l}{and the mass of the core and mantle.}\\
\end{tabular}
\end{table}

\subsection{Earth's missing mass.}

The compositions of Earth and Venus are nearly identical, as shown in Table~\ref{EaVe}.  The total mass of Venus, calculated assuming the same densities as Earth's crust, mantle, and core, is statistically consistent with its measured mass.  The mass of Earth's crust and mantle is 67.6(3) wt\% of its total mass while the mass of Venus' crust and mantle is 71.4(13) wt\% of its total mass.  If both the Earth and Venus initially had the same mass fractions of crust and mantle, then Earth must have lost ${\approx}$4\% of its crust and mantle which would correspond to $2.54(4){\times}10^{11}$ Gt.  This is slightly more than Theia transferred to Mars during the giant impact.  The giant impact also removed crust and mantle to form the Moon, whose mass is $7.35\times10^{10}$ Gt, so correcting for this indicates that $1.80(4)\times10^{11}$ Gt of crust and mantle must have been lost to Mars.  This is nearly identical to the excess mass of crust and mantle that are found on Mars.  Earth's density, 5.51 g/cm$^3$, is significantly higher than Venus' density, 5.24 g/cm$^3$.  Correcting the density for the mass lost to the Moon and Mars reduces to Earth's density to 5.28 g/cm$^3$ and its iron content to 30.8 wt\% which is in better agreement with that of Venus.  

The impact of Theia with Earth is quantitatively confirmed by the mass transfer of crust and mantle to Mars and the Moon.  The composition of Earth before the giant impact is shown to be nearly identical to that of Venus and is consistent with both planets being formed from the same meteoritic material.

\subsection{Earth/Theia collision dynamics}

Both momentum and energy must be conserved during the giant impact, as shown in Eq.~\ref{Bounce}, where $v_1$ is Theia's initial relative velocity,
\begin{equation}
\label{Bounce}
\begin{aligned}
m_1v_1+M_1u_1 =&m_2v_2+M_2u_2\\
\frac{1}{2}m_1v_1^2+\frac{1}{2}M_1u_1^2 =&\frac{1}{2}m_2^2v_2+\frac{1}{2}M_2u_2^2
\end{aligned}
\end{equation}
$v_2$ is Mars final relative velocity, and $u_1$ and $u_2$ are the initial and final velocities of Earth.  The mass $m_1{=}4.5{\times}10^{11}$ Gt is Theia's mass, $m_2{=}6.42{\times}10^{11}$ Gt is Mars mass, $M_1{=}6.42{\times}10^{12}$ Gt is the mass of Earth before the impact, and $M_2{=}5.97{\times}10^{12}$ Gt is the current mass of Earth.  $M_1$ includes the mass of the Moon, $7.35{\times}10^{10}$ Gt, and the mass transferred from Earth to Mars, $m_2{-}m_1{=}1.9{\times}10^{11}$ Gt.  The rebounding Theia would have performed a Hohmann transfer, escaping into the orbit of Mars as shown in Fig~\ref{Horbit} while Earth would gave gone into its current orbit.

Theia's total velocity when it struck the Earth would have been $v_1{=}u_1{+}v_0$, where $v_0=9.35$ km/s is the velocity gained by gravitational attraction when Theia's surface contacted Earth surface as given by Eq.~\ref{V0} where $R_E${=}6378.1 km is the radius of the Earth.
\begin{equation}\label{V0}
\begin{aligned}
E_0=&GM_1m_1/(R_E+R_T)\\
v_0=&\sqrt{2E_0/m_1}
\end{aligned}
\end{equation}
\vspace{-0.4cm}
From Eq.~\ref{RP}, $R_T{=}2740$ km is the radius of Theia calculated assuming
\begin{equation}\label{RP}
R_T=m_1/(4\pi\rho_T/3)^{1/3}
\end{equation}
that Theia had the same density as Venus.
 
\begin{figure}
  \centering
  \includegraphics[width=0.5\textwidth]{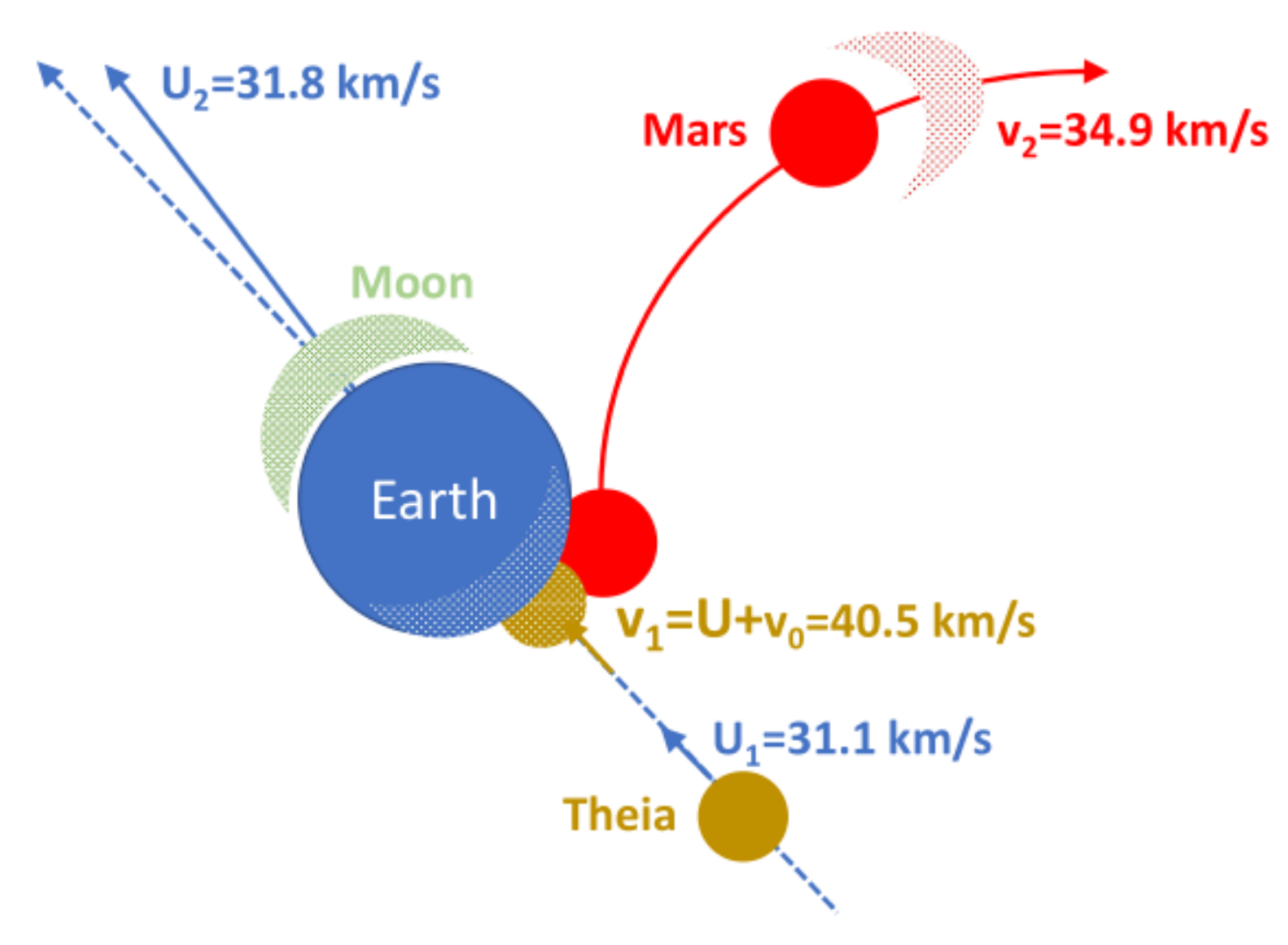}
  \caption{The giant impact of Theia with Earth (not to scale).  Theia initially traveling at the same orbital velocity as Earth gained additional velocity from gravitational attraction before striking Earth.  Both Theia's and Earth's mantle and crust were melted at impact.  As Mars was ejected from the impact it carried away a large amount of Earth's crust and mantle before undergoing a Hohmann transfer to an orbit 1.54 AU from the Sun.  Earth recoiled undergoing a Hohmann transfer to an orbit 1.00 AU from the Sun.  At the antipode of the impact Earth's crust and mantel were ejected into orbit becoming the Moon.}
  \label{EarthMars}
\end{figure}

There are many possible solutions to Eq.~\ref{Bounce} depending on the initial and final velocities of Earth and Theia.  We can select $u_1$ and $u_2$ and solve for $v_2$ with Eq.~\ref{Bounce} and generate values that would support a Hohmann transfer of both Earth and Mars to their current orbits as described In Eq.~\ref{Hohmann} where $r_1$ is the initial distance from the Sun, $r_2$
\begin{equation}\label{Hohmann}
  \Delta v = \sqrt{\frac{GM_{\odot}}{r_1}} \biggl[\sqrt{\frac{2r_2}{r_1+r_2}}-1\biggr]
\end{equation}
is the final distance from the Sun, and $\Delta v =v_2-u_1$ is the impulse velocity.  We can assume that Earth and Theia both initially orbited at a distance $r_1$=0.91 AU from the Sun with a velocity $u_1$=31.12 km/s, and Theia's impact velocity was $v_1$=40.46 km/s.  If Earth recoiled with a velocity 31.82 km/s then the conservation of energy and momentum requires that Mars recoiled from the impact with a velocity of $v_2${=}34.8 km/s as shown in Fig.~\ref{EarthMars}.  For Earth, $\Delta v_E=u_2{-}u_1$=0.69 km/s giving a Hohmann transfer to $r_2$=1.00 AU, its current distance from the Sun.  For Mars, $\Delta v_M = v_2{-}u_1$ = 3.75 km/s giving a Hohmann transfer to $r_2=1.54$ AU, Mars average distance from the Sun today.  This is consistent with both Earth and Mars settling into their current orbits.

\begin{figure}
  \centering
  \includegraphics[width=0.5\textwidth]{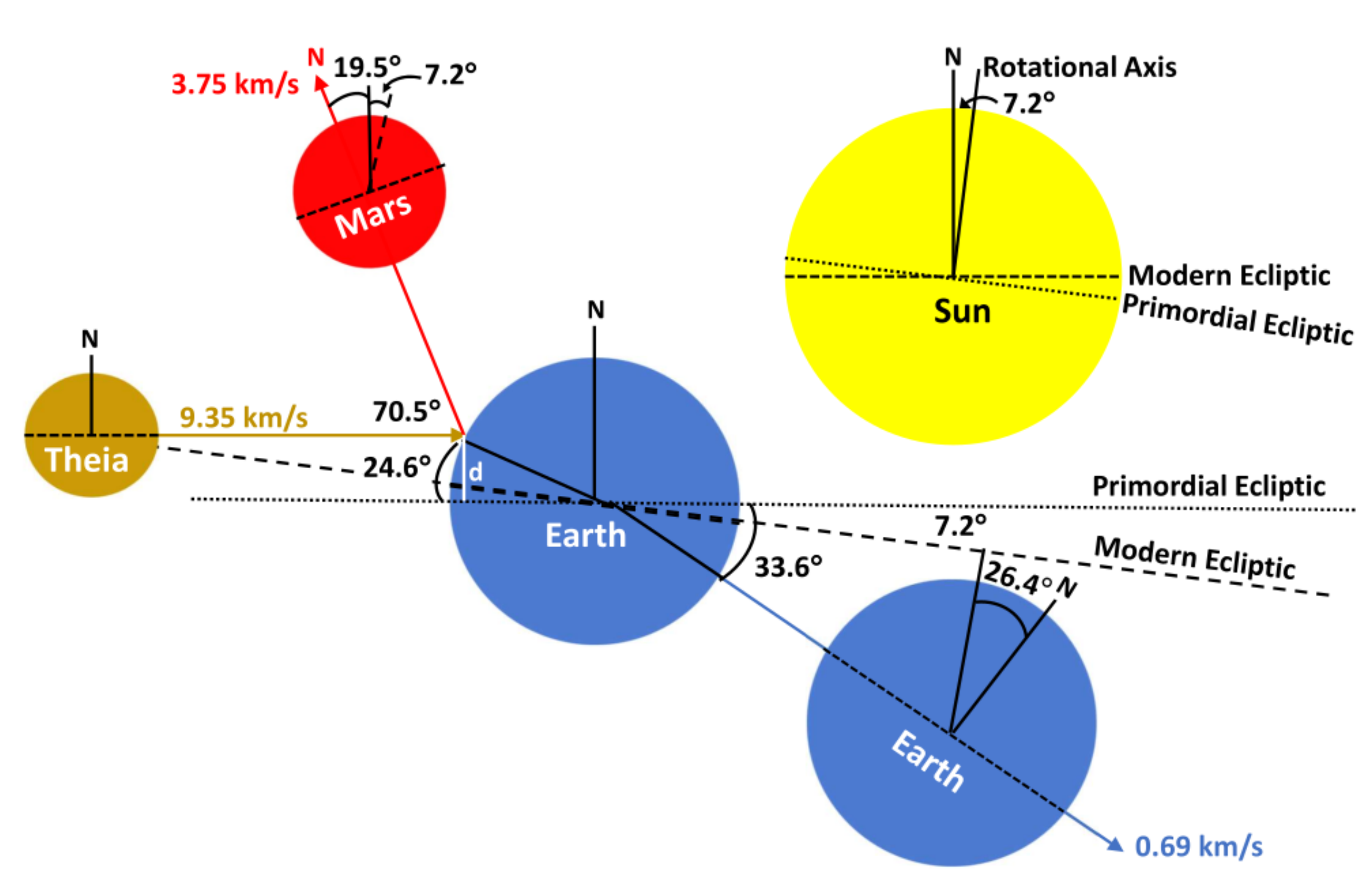}
\caption{Ellipticity of Earth and Mars following the giant impact assuming an impact angle of 24.6$^{\circ}$.  Conservation of momentum requires that the Earth was deflected by 33.6$^{\circ}$ by the impact while Mars rebounded at an angle of 70.5$^{\circ}$ relative to the primordial eclyptic of the Sun's axis.  The Sun's ecliptic was subsequently shifted by 7.2$^{\circ}$ leading to the current axial tilts of Earth and Mars.}
  \label{MAngle}
\end{figure}

\subsection{Rotation and obliquity}

The giant impact would have substantially increased both the rotation rate and obliquity of Mars and Earth.  The rotational frequency depends on the impact parameter $d$, as described in Fig.~\ref{MAngle}, and can be calculated by Eq.~\ref{Omega} where $m_1$ is Theia's mass, $v_0$ is its velocity,
\begin{equation}\label{Omega}
\omega = \frac{m_1 v_0 d}{I_E}
\end{equation}
as described above, $I_E$ is Earth's moment of inertia, and $d$ is the distance above or below the orbital plane where the impact occured.  The impact parameter is not known, a priori, but should be consistent with the observed rotation rates of Mars and Earth.  If we assume an impact angle $\chi{=}24.6^{\circ}$, as shown in Fig.~\ref{MAngle}, then for Earth $d_E{=}R_E \sin(\chi)$=2657 km and $\omega{=}9.62\times10^{-5}$ radians/s resulting in an 18.1 hr day.  The same impact parameter would apply to Theia giving $d_M=R_T \sin(\chi)${=}1412 km corresponding $\omega=7.27\times10^{-5}$ radians/s resulting in an 24.0 hr day, the same as today.  Lacking the tidal effects of a moon, Mars period of rotation should have changed little since it was formed.

The off-center impact induced a large obliquity into the orbits of Earth and Mars.  The angles that Earth and Mars left the giant impact are constrained by the conservation of momentum as given by Eq.~\ref{Angle} where $\theta$ is the angle that Earth departed and $\phi$ is the angle that Mars departed from the impact, as shown in Fig.~\ref{MAngle}.
\begin{equation}
\label{Angle}
\begin{aligned}
  M_2 \Delta v_E sin (\theta)- m_2 \Delta v_M sin (\phi) &= 0\\
  M_2 \Delta v_E cos (\theta)+ m_2 \Delta v_M cos (\phi) &= m_1 v_0\\
  sin (\phi)^2 +cos (\phi)^2 &=1
\end{aligned}
\end{equation}
It is postulated that at the time of the giant impact the Sun's axis of rotation had not yet tilted so the calculated angle $\theta$ and $\phi$ should be considered in the primordial ecliptic.  The Earth was shifted by $\theta=33.6^{\circ}$ which, renormalized to the modern ecliptic as shown in Fig.~\ref{MAngle}, became $\theta=26.4^{\circ}$, Earth's primordial obliquity.  This is slightly higher than Earth's modern obliquity $23.4^{\circ}$.  Mars rebounded by an angle $70.5^{\circ}$ but, as shown in Fig.~\ref{MAngle}, it rebounded elastically maintaining the relative direction of its North pole with respect to its equator.  Correcting for change in the elliptic, Mars obliquity became $26.7^{\circ}$, similar to Earth but slightly higher than its modern value of $25.2^{\circ}$.  Historical Variations in Earth and Mars obliquity is well know.  It was proposed that $\approx$430 million years ago Earth's obliquity was $26.4\pm2.1^{\circ}$~\cite{Williams93}, similar to its primordial value.  Mars obliquity has changed more significantly into recent times ranging from $11-49^{\circ}$ during the past 80 million years~\cite{Touma93}.

If the initial impact angle were $24.6^{\circ}$ North of Earth's equator then, correcting for Earth' modern tilt, the impact latitude should be $1.2^{\circ}$ North today.  This is very close to the $5^{\circ}$ latitude of the Indian Ocean gravitational anomaly.  Conversely, the corresponding impact angle on Mars would have been $24.6^{\circ}$ South so, correcting for Mars' tilt, it should lie at $49.8^{\circ}$ South today.  That is very near the Hellas Basin which is $42^{\circ}$  South.  The assumed impact angle is thus consistent with Earth and Mars rotation rates, obliquities, and the geography of the impact sites.

\section{The origin of the Moon}

The iron abundance of the Moon is $9\pm4$\%~\cite{Parkin73} which is consistent with ${\approx}6\%$ that would be expected if it were composed of ejected crust and mantle.  Earth's crust is widely assumed to have floated out of a magma ocean very shortly after the Earth formed~\cite{Elkins11}.  When this crust and mantle was stripped away from Earth there would have been no magna ocean from which the crust could reform. The thin swath of crust stripped away by the impact should have left a permanent scar on the Earth.  The amount of crust transferred from Earth should be found on the Moon.

\subsection{Earth's missing crust}

The Earth is unique among the terrestrial planets because it has two distinct kinds of crusts.  The continental crust is ancient, dating to the early formation of the Earth, with an average thickness of 34.4$\pm$4.1 km~\cite{Huang13}.  The ocean has a young, thin crust, 8.0$\pm$2.7 km thick~\cite{Huang13}, that is continuously erupting from the ocean trenches and subducting back into the Earth at the boundaries of continental crust.  As shown in Fig.~\ref{MarsCD}, Earth's oceans correspond geographically to Mars' lowlands.  The oceanic crust spans 59\% of the Earth’s surface~\cite{Cogley84}.  This is very different from Venus which has a single continuous, continental-like crust, locked in place over its entire surface.  Volcanism has occurred randomly across the surface of Venus creating broad lava plains, while volcanism on Earth occurs primarily at the edges of the continental plates.  The thin oceanic crust has freed Earth's continental crust to move making plate tectonics possible.

The Earth, like Venus, originally had only a continuous continental crust until the giant impact ejected a swath of Earth's crust and mantle into space.  The lost crust never regenerated leaving a broad swath of exposed magma that became the oceanic crust.  The oceanic crust is the lowland of Earth, like that of Mars, both of which became covered by an ocean.  The mass of continental crust that disappeared from the Earth should now be present on the Moon.  Assuming that the Earth lost 59\% of its 34.4 km thick crust, with a density of 2.83 g/cm$^3$~\cite{Christensen95}, then $(2.9\pm0.3){\times}10^{10}$ Gt of crust must have transferred to the Moon.

The density of the Moon is 3.34 g/cm$^3$, greater than Earth's crust, but less than Earth's upper Mantle, $3.59\pm0.07$ g/cm$^3$~\cite{Nolet11} assuming a 2\% uncertainty.  If the Moon is a mixture of the Earth's crust and upper mantle with no significant core, its density would be consistent with a composition that is 33\% Earth-like crust and 67\% Earth-like mantle.  The mass of the Moon is $7.35{\times}10^{10}$ Gt, so the amount of Earth-like crust on the Moon would be $(2.4\pm0.5){\times}10^{10}$ Gt, consistent with that lost from Earth.  The weighted average mass of crust transferred from the Earth to the Moon is $(2.8\pm0.3){\times}10^{10}$ Gt, which is greater than the current mass of Earth's continental crust, $2.02{\times}10^{10}$ Gt~\cite{Peterson07}.

\subsection{Earth's missing atmosphere}

If the Earth lost 59\% of its crust when the Moon formed, then it should also have lost a comparable amount of its atmosphere.  It has been estimated that the giant impact removed 50\% of Earth's atmosphere~\cite{Ahrens04}.  This can be tested by comparing the carbon and nitrogen abundances on Venus and Earth assuming the initial conditions on both planets were the same.  On Venus nearly all of the surface carbon is found as CO$_2$ in the atmosphere with only a small abundance of N$_2$.  The amount of carbon stored in Earth's crust, ocean and atmosphere is estimated at $5.4{\times}10^7$ Gt, while the amount of carbon in Venus' atmosphere is $1.25{\times}10^8$ Gt~\cite{Lecuyer00}.  Assuming that both Earth and Venus initially had comparable CO$_2$-rich atmospheres and the CO$_2$ abundances are proportional to the planets surface areas then, since Venus' surface area is 90\% that of Earth, the Earth must have lost 61\% of its CO$_2$ following the giant impact.  The remaining CO$_2$ was later transferred mostly into carbonate rock by biological action.  Similarly, the mass of N$_2$ in Earth's atmosphere is $2.1{\times}10^6$ Gt while the mass of N$_2$ in Venus' atmosphere is $4.8{\times}10^6$ Gt~\cite{Lecuyer00}, indicating that the Earth also also lost 61\% of its N$_2$ after the giant impact.

Although the loss of atmosphere on Earth correlates well with the loss of continental crust, this does not rule out later events that may have removed significant amounts of atmosphere. Argon, unlike carbon and nitrogen whose abundances were established when Earth and Venus formed, is continuously injected into the atmosphere because it is constantly produced by the decay of $^{40}K$ in the crust and mantle~\cite{Bender08}.  At the time of the giant impact there was relatively little argon in the atmospheres of Venus and Earth.  Comparison of the modern argon abundances on Earth and Venus is a test on whether additional events may have impacted the atmospheres of either planet after the Moon had formed.  The mass of Earth's atmosphere is $5.15\times10^6$ Gt with an argon abundance of 0.67 wt\%, so the mass of atmospheric argon is $3.47\times10^4$ Gt.  On Venus the atmospheric argon concentration is 70 ppm~\cite{Taylor14} corresponding to a mass of $2.90\times10^4$ Gt.  Correcting for the mass of Venus mass which is 81.6\% that of Earth, the relative argon abundance becomes $3.55\times10^4$ Gt, slightly higher than that on Earth.

The difference between the argon abundance on Venus and Earth is 800 Gt which could be accounted for by the loss of argon during the giant impact.  The current rate of argon outgassing from the Earth's crust and mantle $(4.4\pm0.4){\times}10^6$ kg/yr~\cite{Bender08}, but at the time of the giant impact it would have been 11.5${\times}$ larger due to a larger $^{40}K$ abundance.  Assuming that 61\% of the original argon abundance was lost, 1310 Gt of argon was present in Earth's atmosphere at the time of the giant impact.  That would have taken 26 million years to accumulate after Earth formed.  The age of the Moon has been determined as 4.51 Gyr~\cite{Thiemens19,Barboni17}, which when compared with the age of the Earth, $4.54\pm0.05$ Gyr, suggests that the Moon formed $30\pm5$ million years after the Earth in good agreement with the observed argon deficit.

\subsection{Earth/Moon dynamics}

\begin{figure}
  \centering
  \includegraphics[width=0.4\textwidth]{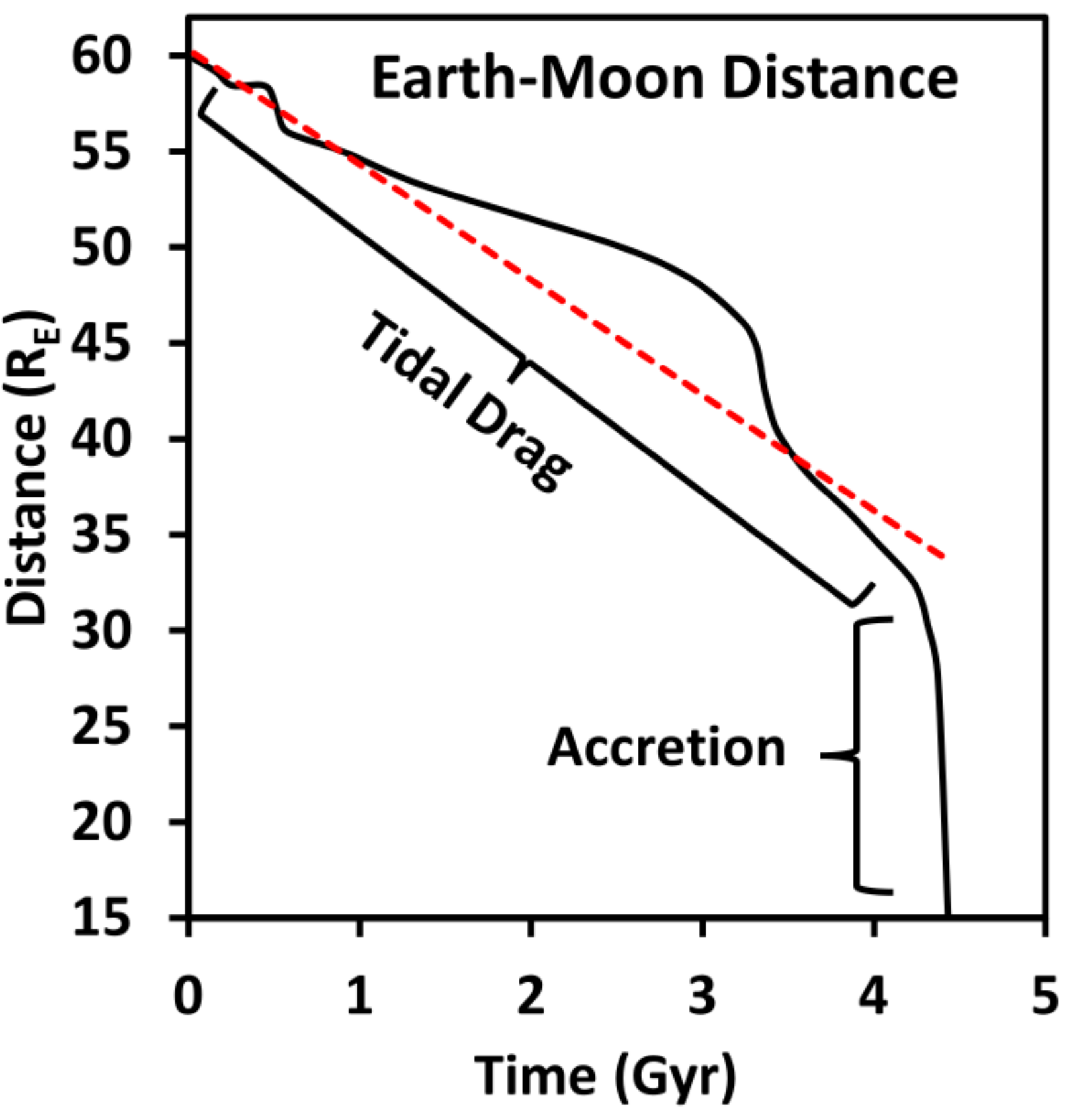}
  \caption{Variation of the distance between the Earth and the Moon since its formation~\protect\cite{Farhat22}.  The dotted rid line is an extrapolation of the Moon's current rate it is retreating from the Earth of 3.83 cm per yr~\protect\cite{Williams16}.}
  \label{E-M}
\end{figure}

The motions of the Earth and the Moon are coupled.  As the Moon has receded away from the Earth, the Earth's rotational rate has slowed.  The energy of the Earth moon system is given by Eq.~\ref{EEM} where $M_E$ 
\begin{equation}\label{EEM}
  E_{tot} = E_{Moon} +E_{Earth} = -\frac{GM_EM_M}{2r_{EM}} + \frac{1}{2}I_E\omega_E^2
\end{equation}
is the mass of the Earth, $M_M$ is the mass of the Moon, $r_{EM}$ is the average distance of the Moon from the earth, $I_E$ is Earth's moment of inertia, and $\omega_E$ is Earth's rotational frequency.  Today Earth's rotational energy is $E_{Earth}=2.14\times10^{29}$ J and, assuming the average distance of the Moon from the Earth is 382,500 km, $E_{Moon}=-3.83\times10^{28}$ J, so $E_{tot}=1.76\times10^{29}$ J.  The total energy of the Earth/Moon system should not have changed since the Moon formed.  If the Earth's rotational frequency were $9.62/times10^{-5}$ radians/s after the giant impact, then its rotational energy was $3.72\times10^{29}$ J implying that $E_{Moon}=-1.96\times10^{29}$ J.  This suggests that the Moon formed 75,000 km (11.7 Earth radii) from Earth.  This is consistent with a model of the historical Earth-Moon distance based on geological evidence~\cite{Farhat22}, as shown in Fig.~\ref{E-M}, where the Earth-Moon distance increased rapidly from $<$15 to $\approx$30 Earth radii in only 100 million years after the giant impact.  During the past 4.5 Gy the Moon has slowly retreated at a rate of 3.83 cm/yr~\cite{Williams16} from 30 to 60 Earth radii due to tidal friction between the Earth and the Moon. 
The early rapid movement of the Moon away from the Earth can be ascribed to the aggregation of Earth's crust and mantle as the Moon quickly gained mass slowing its orbital velocity.

\section{The formation of Mercury}

The planet Mercury has a highly elliptical orbit, 0.377 AU from the Sun, with a composition very different from that of Earth or Venus.  Its iron abundance is 68.47 wt\% more than twice that of either planet, suggesting that if it once had the same iron abundance as Venus it must subsequently have lost a substantial amount of its crust, mantle, and core.  Such a loss has been ascribed to a large impact event~\cite{Benz88}, but the fate of this lost mass is unknown and the dynamics of such an impact make it seem unlikely~\cite{Franco22}.  Another explanation, proposed here, is that Mercury passed close to either Earth or Venus and was slingshot towards the Sun where much of its crust, mantle, and core were stripped away and deposited into the Sun's convection zone where should remain today.  If Mercury's initial iron abundance were 31.2 wt\%, like Venus~\cite{Morgan80}, then its original mass would have been at least $7.31{\times}10^{11}$ Gt, assuming that no core was lost.

\subsection{Convection zone metalicity}

As Mercury passed inside the Sun's convection zone, it would have lost a substantial amount of mass that should still remain trapped there today.  This would have dramatically increased the metalicity, $z$, of the Sun's convective zone with respect to the Sun as a whole.  A discrepancy in spectroscopic measurements of the Sun's metalicity and from dynamic models of the Sun has been observed that is referred to as the "solar modeling problem"~\cite{Vagnozzi19}. Direct measurements give $z{=}0.0202$~\cite{Anders89}, $z{=}0.0170$~\cite{Grevesse98}, and $z{=}0.0196$~\cite{Steiger15} for an average $z{=}0.0189(17)$.  Two solar models are in good agreement giving $z{=}0.0134$~\cite{Asplund09} and $z{=}0.0133$~\cite{Lodders09}, respectively.  A simple explanation for this discrepancy is that direct measurements determine only the metalicity of the Sun's convection zone while the models reflect the expected metalicity of the Sun as a whole.  This is consistent with the high metalicity convective zone being due to the trapping of Mercury's missing mass.  The difference between the direct measurements and the models is $\Delta z=0.0055(17)$ and would correspond to the amount of Mercury's original mass that remains in the convection zone today.

The Sun's radius is $6.9566(14){\times}10^5$ km~\cite{Haberreiter08}, the depth of its convective zone is $1.997(21){\times}10^5$ km~\cite{Christensen91}, and its average density has been reported as $\rho_{conv}=0.104~g/cm^3$~\cite{Bohm67} although that is still considered to be poorly known.  The mass of the Sun's convective zone is thus $M_{conv}=9.35{\times}10^{13}$ Gt and, assuming $\Delta z=0.0055(17)$, the mass of Mercury remaining in the convective zone is $M_{conv}{\times}\Delta z=5.1{\times}10^{11}$ Gt.  The actual mass Mercury lost should be much higher because the solar wind has expelled much of Mercury's original lost mass.

\begin{table*}
  \centering
  \small
\tabcolsep=2pt
 \resizebox{16.5cm}{!}{
\begin{tabular}{lrlrl|lllll|ccccc}
\toprule
&&&&\multicolumn{1}{c}{}&&&\multicolumn{1}{c}{}&\multicolumn{1}{c}{}&\multicolumn{1}{c}{Galactic}&Mercury&Galactic&Mercury&Solar\\
&\multicolumn{2}{c}{Solar}&\multicolumn{2}{c}{Galactic}&\multicolumn{2}{c}{Mercury Crust$^a$}&\multicolumn{2}{c}{Mercury Core$^a$}&\multicolumn{1}{c}{+Mercury}&/Solar&/Solar&/Solar&wind\\
&ppm&\multicolumn{1}{c}{Mass (Gt)}&ppm&\multicolumn{1}{c}{Mass (Gt)}&\multicolumn{1}{c}{wt\%}&\multicolumn{1}{c}{Mass
(Gt)}&wt\%&\multicolumn{1}{c}{Mass (Gt)}&\multicolumn{1}{c}{Mass (Gt)}&Ratio&Ratio&Mass Ratio&Correction$^b$\\
\midrule
C &3900&$3.7\times10^{11}$&3300&$3.0\times10^{11}$&0.0011&$2.9\times10^{8}$&0.8&$2.9\times10^{8}$&$3.1\times10^{11}$&322&1.02&0.83&\\
N&1100&$1.0\times10^{11}$&700&$6.4\times10^{10}$&0.00020&$4.3\times10^{6}$&0.009&$3.3\times10^{6}$&$6.4\times10^{10}$&4569&0.76&0.62&\\
O&9100&$8.5\times10^{11}$&7400&$6.9\times10^{11}$&44.3&$2.1\times10^{11}$&2.7&$9.5\times10^{8}$&$9.0\times10^{11}$&=1.00&=1.00&1.06&1.06\\
Na&41&$3.9\times10^{9}$&20&$1.3\times10^{9}$&0.26&$1.3\times10^{9}$&0.14&$4.9\times10^{7}$&$2.6\times10^{9}$&0.32&0.42&0.67&\\
Mg&980&$9.2\times10^{10}$&580&$3.8\times10^{10}$&22.3&$1.1\times10^{11}$&0.59&$2.1\times10^{8}$&$1.4\times10^{11}$&0.11&0.52&1.58&1.17\\
Al&81&$7.5\times10^{9}$&50&$3.3\times10^{9}$&2.29&$1.1\times10^{10}$&&&$1.4\times10^{10}$&0.09&0.54&1.88&\\
Si&1010&$9.4\times10^{10}$&650&$4.3\times10^{10}$&21.3&$1.0\times10^{11}$&5.0&$1.8\times10^{9}$&$1.5\times10^{11}$&0.13&0.56&1.55&1.01\\
P&9&$8.3\times10^{8}$&7&$4.6\times10^{8}$&0.008&$1.4\times10^{8}$&0.28&$9.9\times10^{7}$&$6.0\times10^{8}$&1.03&0.69&0.73&\\
S&500&$4.7\times10^{10}$&440&$2.9\times10^{10}$&0.024&$7.6\times10^{8}$&1.8&$6.5\times10^{8}$&$3.0\times10^{10}$&11.72&0.77&0.64&\\
K&5&$4.5\times10^{8}$&3&$2.0\times10^{8}$&0.024&$1.2\times10^{8}$&0.021&$7.4\times10^{6}$&$3.2\times10^{8}$&0.51&0.55&0.72&\\
Ca&93&$8.7\times10^{9}$&60&$4.0\times10^{9}$&2.5&$1.2\times10^{10}$&&&$1.6\times10^{10}$&0.10&0.57&1.81&\\
Ti&5&$4.2\times10^{8}$&3&$2.0\times10^{8}$&0.12&$5.6\times10^{8}$&&&$7.6\times10^{8}$&0.11&0.58&1.79&\\
Cr&24&$2.3\times10^{9}$&15&$9.9\times10^{8}$&0.26&$1.5\times10^{9}$&0.66&$2.4\times10^{8}$&$2.5\times10^{9}$&0.21&0.55&1.09&\\
Mn&20&$1.9\times10^{9}$&8&$5.3\times10^{8}$&0.10&$6.1\times10^{8}$&0.33&$1.2\times10^{8}$&$1.1\times10^{9}$&0.27&0.35&0.60&\\
Fe&1900&$1.8\times10^{11}$&1100&$7.2\times10^{10}$&6.3&$6.0\times10^{10}$&82.8&$3.0\times10^{10}$&$1.3\times10^{11}$&0.37&0.51&0.75&1.20\\
Ni&103&$9.7\times10^{9}$&60&$4.0\times10^{9}$&0.19&$2.7\times10^{9}$&5.1&$1.8\times10^{9}$&$6.7\times10^{9}$&0.45&0.51&0.69&\\
\hline
\rule{0pt}{3ex}
Sum&18900&$1.8\times10^{12}$&13400&$1.3\times10^{12}$&100&$4.8\times10^{11}$&100&$3.6\times10^{10}$&$1.8\times10^{12}$&Average&0.62&1.1(5)&1.11(9)\\
\bottomrule
\multicolumn{14}{l}{$^a$ From ~\cite{Wang18}, calculated from elemental abundances assuming Mercury lost $5.1{\times}10^{11}$ Gt of crust, mantle and core.}\\
\multicolumn{14}{l}{$^b$Ratio corrected for solar wind abundance ratio O/Mg/Si/Fe=1.00/0.14/0.17/0.13~\cite{Pilleri15,Reisenfeld13}}\\
\end{tabular}}
  \caption{Elemental abundance in the Sun's convection zone.  The observed elemental abundances~\protect\cite{Lodders21}, assuming a metalicity of 0.0189, are compared with the sum of the Galactic abundance~\protect\cite{Croswell96}, assuming a metalicity of 0.0134,  and the sum of crustal, mantel, and core mass lost by Mercury during its passage through the Sun's convection zone. }
  \label{Mercury}
\end{table*}
The relative elemental abundances in the convective zone should reflect contributions from a primordial galactic component~\cite{Croswell96}, with a metalicity of $z=0.0134$, and an Earth-like terrestrial component~\cite{Wang18}, from Mercury.  The galactic elemental abundances, as shown in Table~\ref{Mercury}, cannot alone explain the solar elemental abundances unless the Sun were substantially enriched in oxygen.  Combining the galactic and terrestrial elemental abundances gives a better agreement with the convective zone abundances with an average elemental ratio of (Galactic+Mercury)/Solar=1.1(5) albeit with significant fluctuations.  The component of Mercury's mass found in the Sun's convective zone should be corrected for losses due to the solar wind during the history of the Sun.  As the convection zone loses mass to the solar wind it is replaced by mass rising from the Sun's interior diluting Mercury's contribution to the metalicity.  The Sun is estimated to have lost a range of $10^{-12}-10^{-14}~M_{\odot}$ per year to the solar wind during its lifetime.  Assuming an average rate of $10^{-13}~M_{\odot}/yr$, the convection zone would have lost $4{\times}10^{14}$ Gt, or ${\approx}10\times$ its current mass. If a proportional amount of Mercury's original lost mass also disappeared into the solar wind, then its initial mass would have been ${\approx}5\times10^{12}$ Gt, comparable to the masses of Earth and Venus, implying that Mercury actually lost 94\% of its total mass as it passed through the Sun's convective zone.

The relative abundances of the primary elements O, Mg, Si, and Fe in the solar wind are different from the relative abundances in the convective zone.  The current metalicity of the convective zone must be corrected for the differential elemental mass loss due to the solar wind over time.  The solar wind is poorer in Fe and richer in Mg and Si than the convective zone.  Correcting the abundance ratio in Table~\ref{Mercury} for the solar wind losses gives (Galactic+Mercury)/Solar-1.11(9), in excellent agreement with the hypothesis that much of Mercury's original mass was captured into the Sun's convection zone.

\subsection{The mystery of graphite on Mercury's surface}

The surface of Mercury is rich in graphite with a carbon content ranging up to 5 wt\%~\cite{Klima18}.  This is commonly described as either having an exogenic in origin, delivered by comets, or an endogenic origin, failing to be absorbed by the core and instead rising to the surface as a graphite flotation crust~\cite{Vander15}.  Both explanations are exotic and were never observed elsewhere in the solar system.  If Mercury passed through the Sun's convective, none of these explanations would be necessary.

An alternate explanation is that Mercury passed so close to the Sun that its CO$_2$-rich atmosphere was carbonized.  Mercury, like all of the terrestrial planets, would originally have had a dense, CO$_2$ rich atmosphere .  The temperature needed to dissociate CO$_2$ to graphite is ${\approx}4800^{\circ}$K~\cite{Kwak15}, significantly higher than temperatures achievable in large impacts~\cite{Steenstra20}, but less than $5800^{\circ}$K at the surface of the convection zone~\cite{Rouan11}.  The existence of large carbon deposits on Mercury's surface is consistent with Mercury passing so close to the Sun that its CO$_2$ atmosphere was decomposed into graphite that remains on its surface today.

\subsection{The slingshot of Mercury}

\begin{figure}
  \centering
\includegraphics[width=0.4\textwidth]{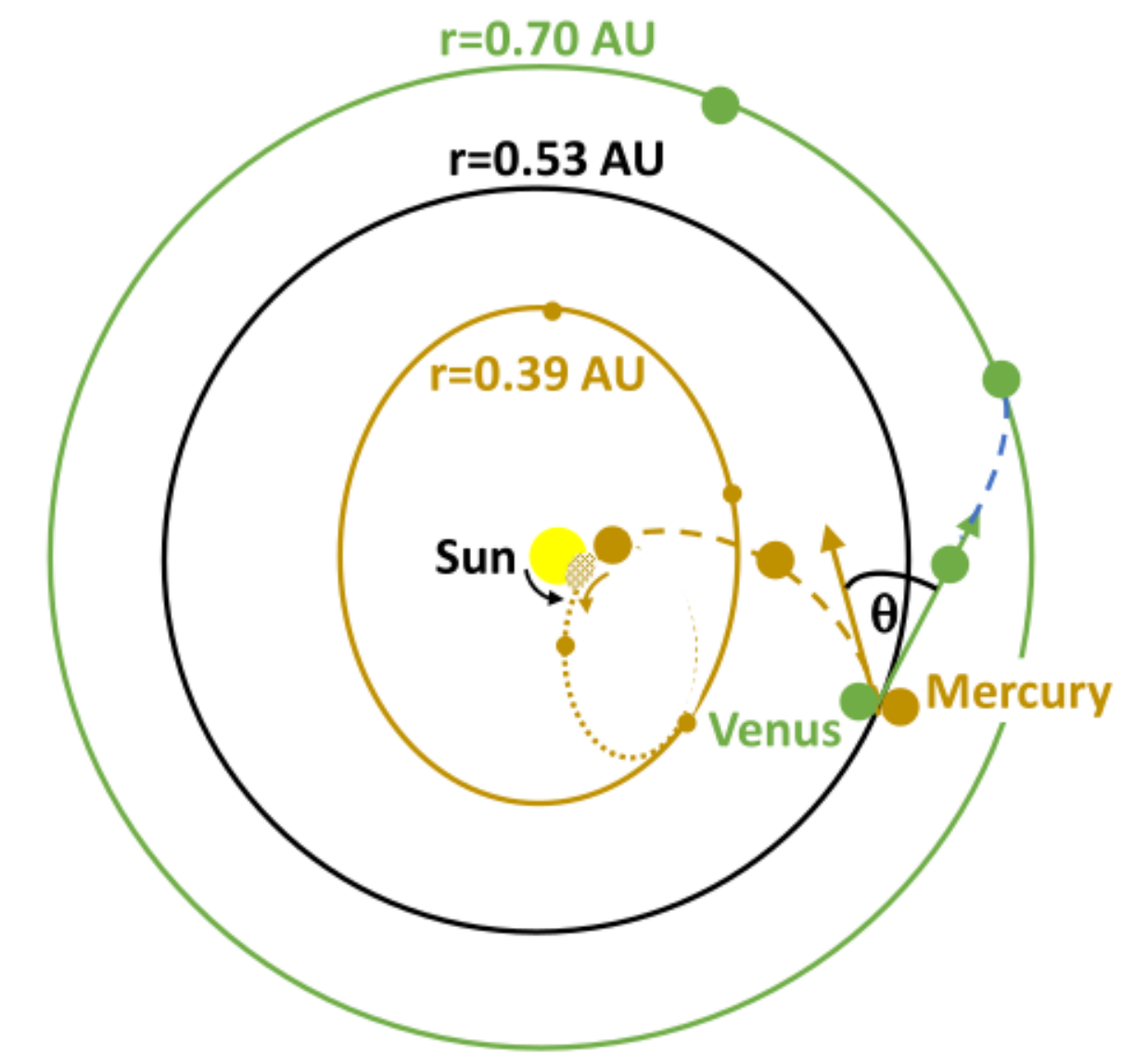}
  \label{Slingshot1}
  \caption{Slingshot of Mercury around Venus and towards the Sun.  Mercury is swung by an angle $\theta$ from its initial orbit reaching terminal velocity at the Sun's convection layer where it loses most of its mass.  It arrives moving in the opposite direction of the Sun's rotation decreasing the rotation rate.  Mercury rebounds at high velocity undergoing a Hohmann transfer orbit into its current orbit.}
  \label{Slingshot}
\end{figure}

The forces that caused Mercury to crash into the Sun are unclear.  It is difficult to send an object to the Sun because of the large energy required.  The Parker Probe sent an object from the Earth the Sun by a slingshot past Venus.   It is proposed here that Mercury and Venus initially orbited near each other and that Mercury was slingshot toward the Sun and Venus towards its current orbit as described in Fig.~\ref{Slingshot}.

The dynamics of the slingshot of Mercury around the planet is described in Eq.~\ref{Sling} where v$_1$ is the Mercury's approaching velocity, v$_2$ is its final velocity, $U$ is the planets orbital velocity, and $\theta$ is the
\begin{equation}
\label{Sling}
v_2 = (v_1-2U)\sqrt{1 + \frac{4Uv_1(1-cos(\theta))}{(v_1-2U)^2}}
\end{equation}
impact angle.  If Mercury initially orbited near Venus it would have gained an additional velocity, $v_0$, from gravitational attraction as it approached Venus, and it would attain a maximum velocity of $v_1=U+v_0$.  Assuming that Mercury came within 100,000 km of Venus, then $v_0$=2.48 km/s.  If Venus and Mercury initially orbited $r_1$=0.53 AU from the Sun and Mercury slingshot at an angle $\theta=20^{\circ}$ towards the Sun, then U=40.9 km/s, $v_1=U+v_0=43.4$ km/s, and $v_2=43.6$ km/s.  Mercury would reach a terminal velocity of 616 km/s as it reached the Sun's convection zone.

Conservation of momentum requires that Venus recoiled with a velocity $U_2$=43.7 km/s.  This would send Venus into a Hohmann transfer orbit~\cite{Hohmann60}, as described by Eq.~\ref{Hohmann}, where $\Delta v=v_2 - U=2.8$ km/s, giving $r_2$=0.70 AU, Venus' current distance from the Sun.  If Mercury made a Hohmann transfer from the edge of the Sun where $r_1$=695,700 km, then for $r_2=5.8{\times}10^7$ km, the average distance of Mercury from the Sun today, then $\Delta v$=177 km/s, a value consistent with Mercury's large impact velocity.  The survival of a large body like Mercury passing close to the Sun is not surprising.  Large comets have often passed through the Sun's convective zone and survived~\cite{Marsden05,Ye14}.  For example, the comet C/2011 W3 (Lovejoy) which was ${\approx}500$ km in diameter, passed within 100,000 km of the Sun's surface and survived~\cite{Sekanina12}.

\subsection{Rotation of the Sun}

When Mercury reached the Sun it was traveling with a velocity of 616 km/s, depositing most of its mass into the convection zone.  This would have produced a large torque changing the Suns rotation rate.  Assuming that its mass was ${\approx}5{\times}10^{12}$ Gt, Mercury would have provided $E_M{\approx}9.5{\times}10^{35}$ J of rotational energy.  The moment of inertia of the Sun is $I_S=6.74{\times}10^{46}$ kgm$^2$, assuming a moment of inertia factor of 0.07, so its rotation rate is $\omega_S=2.97{\times}10^{-6}$ rad/s.  The Sun's equatorial rotation period is 24.47 d, so its rotational energy is $E_S=0.5*I_S*\omega_S^2=2.98{\times}10^{35}$ J.  This is only 32\% of the rotational energy that could be provided by Mercury's impact.  The Sun rotates much more slowly than most theories predict~\cite{Mestel65,Ray12}.  It is estimated that the early Sun rotated with a period of 8 days corresponding to $\omega_S\approx9{\times}10^{-6}$ rad/s~\cite{Mestel65}.  If Mercury applied a torque in the opposite direction of the Sun's rotation it would have slowed the Sun's rotation rate to its current value.  The Sun's initial rotational rate would then be $\omega_S^0=\sqrt{\frac{2(E_M+E_S)}{I_S}}=6{\times}10^{-6}$ rad/s corresponding to a rotation period of 12 d.

\begin{figure}
  \centering
\includegraphics[width=0.5\textwidth]{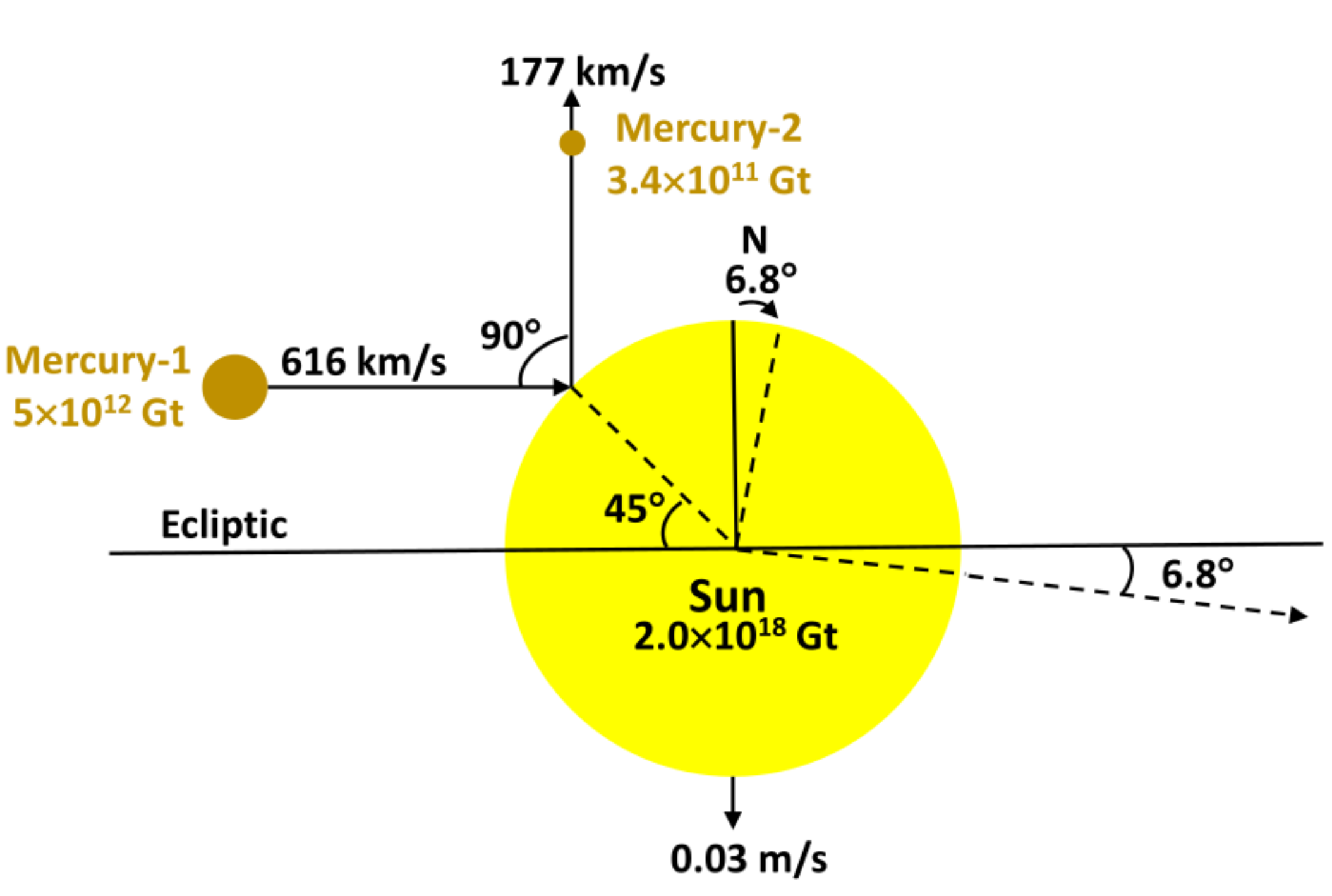}
  \caption{The tilt of the Sun's polar axis is consistent with a collision with Mercury at an impact angle of 45$^{\circ}$ during which Mercury lost 94\% of its mass to the Sun's convection zone.}
  \label{SolarTilt}
\end{figure}

The Sun's axis is tilted by 7.25$^{\circ}$ with respect to the ecliptic.  The origin of this tilt remains a mystery although it has been ascribed to the influence of a giant, distant planet~\cite{Touma93} although no such object is known to exist.  An impact large enough to change the Sun's rotation rate would likely also tilt its axis.  This can be tested by varying the impact angle and requiring the conservation of momentum.  As shown in Fig.~\ref{SolarTilt}, assuming an impact angle of 45$^{\circ}$, Mercury rebounds at an angle of 90$^{\circ}$ providing the maximum possible solar tilt.  If Mercury arrived with a velocity of 616 km/s and departed with a velocity of 177 km/s the conservation of momentum requires that the Sun was tilted by 6.8$^{\circ}$, very nearly its current value.

A caveat to this discussion is that the rotation of the Sun is more complex than that of the planets.  Although the equatorial rotation rate was assumed in these calculations, it varies by up to 50\% with latitude.  The rotation rate of the interior of the Sun may be very different.  Nevertheless, the large kinetic energy provided by Mercury must have had a profound effect on the Sun's rotation and tilt.  More detailed calculations are needed to determine more accurately how the the impact of Mercury changed the Sun's rotation.

\section{The orientation of the terrestrial planets major axes}

\begin{figure}
  \centering
\includegraphics[width=0.4\textwidth]{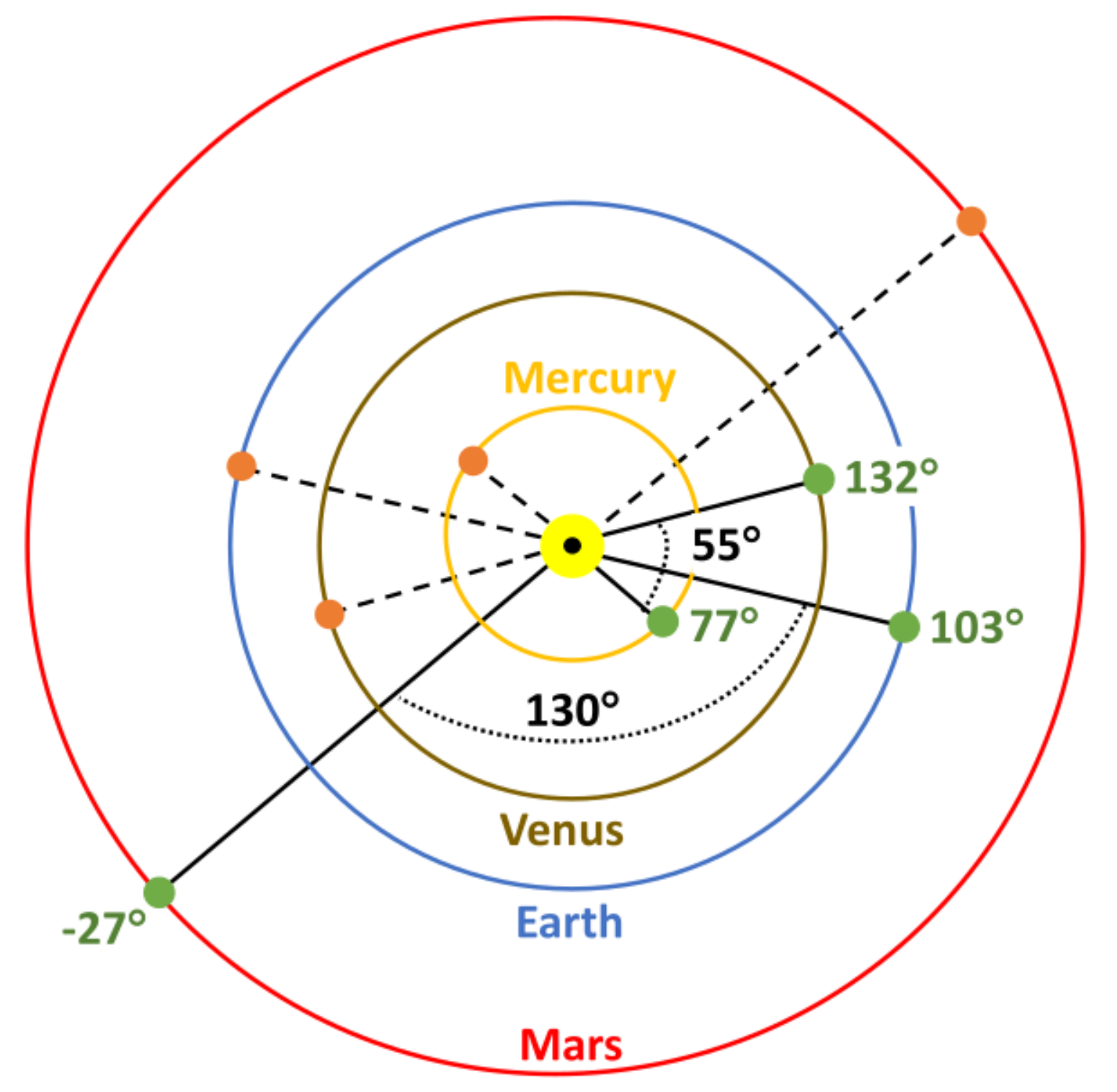}
  \caption{Longitude of perihelion of the terrestrial planet orbits~\protect\cite{Seidelmann92}.  The aphelion (Orange) and perihelion (green) positions for each orbit are indicated.  The major axes of Earth and Mars are separated by 127$^{\circ}$ and the major axes of Venus and Mercury are separated by 55$^{\circ}$.}
  \label{ECC}
\end{figure}

The planets all initially formed in nearby circular orbits~\cite{VanEylen15}.  Collisions between the planets and protoplanets pushed them into new, elliptical orbits.  The longitudes of perihelion for the terrestrial planets are shown in Fig.~\ref{ECC}.
The angle between Mars' and Earth's orbital major axes is $130^{\circ}$, nearly the same as $124^{\circ}$ from the rebound impact calculations.  When Mercury slingshot past Venus, their major axes would have been aligned in opposite directions, $180^{\circ}$ apart.  After Mercury recoiled from the Sun, it would have reversed direction recoiling at an angle of $70^{\circ}$ with respect to Venus assuming the impact were directed towards the center of the Sun as shown in Fig.~\ref{Slingshot}.  Assuming that Mercury struck the sun within $10^{\circ}$ of the center of the Sun, it would have recoiled at an angle between $50-90^{\circ}$ of Venus orbit.  This is consistent with the $55^{\circ}$ angular separation of Venus' and Mercury's major axes today.

\section{Discussion}

Three planets of comparable size, Earth, Venus, and Mercury, and the smaller protoplanet Theia formed out of the solar nebula with nearly identical, meteoritic compositions.  About 30 million years later Theia collided with Earth in a giant impact rebounding as the planet Mars while ejecting Earth's crust and mantle at the impact antipode forming the Moon.  Mars and Earth recoiled by Hohmann transfers to their current orbits.  The impact removed ${\approx}$4\% of Earth's crust and mantle which is quantitatively accounted for on the Moon and Mars.  The giant impact occurred at an impact angle of approximately 24.6$^{\circ}$, tilting the rotational axes of Earth and Mars by 26.4$^{\circ}$ and 26.7$^{\circ}$, respectively.  The impact increased Earth's rotation period to an 18 hr day and Mars' rotation period to a 24 hr day.

The giant impact was centered at the Indian ocean gravitational anomaly on Earth and the Hellas Basin on Mars.  The antipodes of the impact correspond to the Northern lowlands of Mars and the oceans of Earth.  59\% of Earth's crust was ejected near the antipode and was never regenerated leaving Earth with both a terrestrial and an oceanic crust.  The crust lost by Earth is quantitatively accounted for on the Moon.  Simultaneously Earth lost 61\% of its atmosphere.  On Mars the Northern lowlands were similarly covered by a vast ocean, and substantial volcanic activity occurred at the edges of both Earth and Mars oceans.  The loss of crust on Earth allowed plate tectonics to occur as the oceanic crust slid beneath the continental crust.  Plate tectonics likely also occurred on Mars~\cite{Sleep94}.  Water continuously flows from the oceans to the mantle on Earth lubricating continental drift, but as Mars lost its oceans plate tectonics appears to have stopped there.

After the giant impact Mercury approached close to Venus where it was slingshot towards the Sun.  About 94\% of Mercury's mass was incinerated in the Sun's convective zone accounting for its current high metalicity.  The encounter was hot enough to decompose Mercury's CO$_2$-rich atmosphere leaving large deposits of carbon on the surviving planet's surface.  The impact momentum was sufficient to slow the Sun's rotation rate and tilt its axis.  The rebounding Mercury escaped with sufficient velocity to perform a Hohmann transfer into its current orbit.

The ability of life to form on Earth can be ascribed to the giant impact.  All terrestrial planets initially form with dense CO$_2$-rich atmospheres because that is the most stable oxidation state of carbon.  CO$_2$ is a powerful greenhouse gas dooming planets to a hot future devoid of life.  Venus, for example, could only have sustained life for less than 1 Gyr~\cite{Way16,Krissansen21} before becoming the hottest planet in the solar system.  Earth, too, would have had the same fate except that the giant impact removed 61\% of its CO$_2$-rich atmosphere and pushed it farther from the Sun.  Cyanobacteria evolved on Earth $>$3.9 Gyr ago~\cite{Betts18}.  Although its origin remains unclear, it has been proposed that cyanobacteria may have originated in phosphorus-rich hydrothermal plumes on early Earth~\cite{Rasmussen21}.  The giant impact created the ocean basins where volcanic activity was constrained to the continental margins and the ocean trenches where the oceanic crust emerged and subducted.  On Venus there were no hydrothermal vents.  Volcanic activity occurred in the highlands where no oceans could exist, so it is doubtful that life could have formed early on Venus.  The advent of life on Earth quickly removed CO$_2$ from the atmosphere forming stromatolites and ultimately limestone and shale and forever forestalling the runaway greenhouse that befell Venus.

The giant impact also created comparable conditions for life to form on Mars where large oceans formed in the Northern lowlands bordered by volcanic activity.  Although the water disappeared from Mars' surface $\approx3$ Gyr ago~\cite{Kite22}, this would be sufficient time for life to occur there as well.  Indeed, visual evidence is found for stromatolite-like structures at Gusev crater~\cite{Bianciardi15}.  This location appears associated with fumarolic activity, hot springs, and/or geysers~\cite{Ruff20}, comparable to where life may have formed on Earth.  If life formed on Mars it is likely robust and continuing to thrive on Mars below ground.

All of these events are supported rigorously by the mass balance between the Earth, Mars, Moon, Venus, Mercury, and the Sun assuming that the early terrestrial planets all had the same meteoritic composition.  They are also consistent with the conservation of energy and momentum necessary to leave the planets in their modern orbits.  The giant impact is also is also confirmed by geographical evidence linking the locations of the impact sites with the dynamics of the event.  This is a unified, self-consistent theory for how the terrestrial planets formed that has important implications towards our understanding of how other planetary systems may have formed and how the metalicity of stars can be interpreted.

\section{Acknowledgements}

This work was performed under the funding of the University of California retirement system.

\bibliographystyle{unsrtnat}

\end{document}